\documentclass[prl,twocolumn,showpacs,superscriptaddress]{revtex4}

\usepackage{graphicx}
\usepackage{amsmath}
\usepackage{amssymb}
\usepackage{epsfig}

\renewcommand{\v}[1]{{\bf #1}}

\newcommand{\w}{{\omega}}

\def\eqa{\begin{eqnarray}}
\def\eea{\end{eqnarray}}
\newcommand{\eq}{\begin{equation}}
\newcommand{\ee}{\end{equation}}
\newcommand{\nn}{\nonumber\\}
\newcommand{\Eq}[1]{Eq.~(\ref{#1})}

\renewcommand{\Re}{{\rm Re}}
\newcommand{\p}{\partial}

\newcommand{\ua}{\uparrow}
\newcommand{\da}{\downarrow}
\newcommand{\ra}{\rightarrow}

\newcommand{\al}{\alpha}

\newcommand{\del}{\delta}
\newcommand{\Del}{\Delta}
\newcommand{\eps}{\epsilon}

\newcommand{\Ga}{\Gamma}

\newcommand{\La}{\Lambda}

\newcommand{\si}{\sigma}

\newcommand{\nd}{ {\vphantom{\dagger}} }

\usepackage{color}

\begin{document}

\title{Functional renormalization group and variational Monte Carlo studies
of the electronic instabilities in graphene near 1/4 doping}

\author{Wan-Sheng Wang}
\affiliation{National Lab of Solid State Microstructures, Nanjing University,
Nanjing, 210093, China}

\author{Yuan-Yuan Xiang}
\affiliation{National Lab of Solid State Microstructures, Nanjing University,
Nanjing, 210093, China}

\author{Qiang-Hua Wang}
\affiliation{National Lab of Solid State Microstructures, Nanjing University,
Nanjing, 210093, China}

\author{Fa Wang}
\affiliation{Department of Physics, Massachusetts Institute of Technology,
Cambridge, MA 02139, USA}

\author{Fan Yang}
\affiliation{Department of Physics, Beijing Institute of Technology,
Beijing 100081, China}

\author{Dung-Hai Lee}
\affiliation{Department of Physics, University of California at Berkeley,
Berkeley, CA 94720, USA}
\affiliation{Materials Sciences Division, Lawrence Berkeley National Laboratory,
Berkeley, CA 94720, USA}


\begin{abstract}
We study the electronic instabilities of near 1/4 electron doped
graphene using the singular-mode functional renormalization group,
with a self-adaptive k-mesh to improve the treatment of the van
Hove singularities, and variational Monte-Carlo method.  At 1/4
doping the system is a chiral spin density wave state exhibiting
the anomalous quantized Hall effect. When the doping deviates from
1/4, the $d_{x^2-y^2}+i d_{xy}$ Cooper pairing becomes the leading
instability. Our results suggest that near 1/4 electron- or hole-
doping (away from the neutral point) the graphene is either a
Chern insulator or a topoligical superconductor.
\end{abstract}


\maketitle

\section{Introduction}

Graphene, a single atomic layer of graphite, has been a focus of
interest since the pioneering work of Novoselov and
Geim\cite{Novoselov}. At the fundamental level the past research
activities on graphene mostly focused on exploring the
consequences of the unique Dirac-like bandstructure\cite{antonio}.
On the experimental side, few exceptions include the observation
of the fractional quantum Hall effect\cite{eva,kim}, the detection
of the Fermi velocity renormalization\cite{lanzara}, and the
possible observation of ``plasmaron'' in angle-resovled
photosemission\cite{eli}. In general the effects of
electron-electron interaction on the properties of graphene remain
a frontier of this field. Previously based on the
resonating-valence-bond\cite{anderson} concept Pathak {\it et
al.}\cite{baskaran} and Black-Schaffer and Doniach \cite{doniach}
proposed that doped graphene should be a high temperature
superconductor with $d+id'$ pairing symmetry. (Henceforth $d$ and
$d'$ are used to denote interchangeably $d_{x^2-y^2}$ and $d_{xy}$
symmetries, respectively.) In particular, the possibility of
unusual superconductivity and other orders in doped graphene with
van Hove singularities at (or near) the Fermi level becomes a hot
issue.\cite{vanhove,castro} Most recently by a perturbative
renormalization group calculation Nandkishore {\it et al.}
concluded that the van Hove singularities on the Fermi surface
drive chiral $d+id'$ superconductivity in the limit of vanishing
interaction strength\cite{chubukov}.

On a different front Tao Li recently proposed that due to the
existence of Fermi surface nesting 1/4 electron doped Hubbard
model on honeycomb lattice favors the formation of a magnetic
insulating state which possesses nonzero spin chirality and
exhibit the anomalous quantized Hall effect, hence is a Chern
insulator\cite{tao}. Thus near quarter doping graphene suddenly
becomes a playing ground where either a Chern insulator or a
topological superconductor can potentially be realized. Because
the realization of either phases in heavily doped graphene will be
truly exciting, we feel it is meaningful to examine this problem
using the more realistic band structure and interaction
parameters.

In view of the heavy doping we use the Hubbard interaction to
model the screened Coulomb interaction. We perform singular-mode
functional renormalization group (SM-FRG)\cite{husemann} and
variational Monte Carlo (VMC) calculations to address the possible
electronic instabilities. Since the interaction strength is
estimated to be a fraction of the band width, we believe SM-FRG
should yield qualitatively correct answer. The VMC is used to
further confirm such belief. The main results are summarized as
follows. At 1/4 electron doping and for interaction strength
appropriate for graphene we found the chiral spin density wave
(SDW) state is the dominating instability. When the doping level
slightly deviates from 1/4 we find the $d+id'$ pairing instability
surpasses that of the chiral SDW. We propose a schematic phase
diagram in Fig.~\ref{phase-diagram}(b). As in pnictides and
overdoped cuprates\cite{zhai}, the pairing mechanism is due to a
strong scattering channel shared by the SDW and pairing.

\section{Model}
The real-space hamiltonian we used is given by \eqa
H=&&-\sum_{(ij)\si} (c^\dagger_{i\si} t_{ij}^\nd c_{j\si}^\nd
+{\rm h.c.})-\mu N_e+U\sum_i n_{i\ua}n_{i\da}\nn &&
+\frac{1}{2}V\sum_{i\del} n_i n_{i+\del},\eea where $(ij)$ denotes
bonds connecting sites $i$ and $j$,
$\si$ is the spin polarity, $\mu$ is the chemical potential, $N_e$
is the total electron number operator, the $U$-term is the on-site
Hubbard interaction and $V$ is the Coulomb interaction on
nearest-neighbor bonds $\del$. The honeycomb lattice has two
sublattices which we denote as $A$ and $B$ henceforth. As
suggested in Ref.~\cite{antonio} we take $t_1=2.8$ eV, $t_2=0.1$
eV, and $t_3=0.07$ eV for hoppings between the first, second the
third neighbors, respectively, and set $U=3.6 t_1$. As for $V$, we
expect $V<U$ in doped graphene, and take $V=t_1$ as a typical
upper bound. Theoretically, enriched phases may appear for even
larger values of $V/U$.\cite{casten,castro}

\section{Method}

\begin{figure}
\includegraphics[width=8cm]{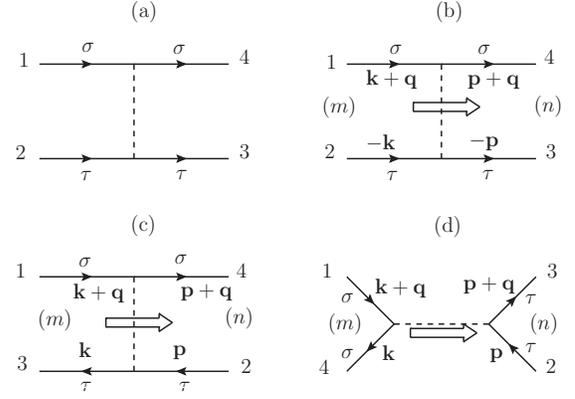}
\caption{A generic 4-point vertex (a) is rearranged into the
pairing (b), crossing (c) and direct (d) channels. Here $\v k,\v
q,\v p$ are momenta, $\si$ and $\tau$ denote spins which are
conserved during fermion propagation, and $m,n$ denote the form
factor (see the text for details). The open arrows indicate
collective propagation.}\label{pcd}
\end{figure}

The SM-FRG method\cite{husemann} we used is a modification of the
FRG method\cite{Salmhofer} applied to the cuprates\cite{honerkamp}
and pnictides\cite{fa}. Fig.\ref{pcd} (a) shows a generic 4-point
vertex function $\Ga_{1234}$ which appears in the interaction
$c^\dagger_1 c^\dagger_2 (-\Ga_{1234}) c_3 c_4$. Here $1,2,3,4$
represent momentum (or real space position) and sublattice. The
spin $\si$ and $\tau$ are conserved during fermion propagation,
and will be suppressed henceforth. Figs.\ref{pcd}(b)-(d) are
rearrangements of (a) into the pairing (P), the crossing (C) and
the direct (D) channels in such a way that a collective momentum
$q$ can be associated and the other momentum dependence can be
decomposed as, \eqa && \Ga_{\v k+\v q,-\v k,-\v p,\v p+\v q}\ra
\sum_{mn}f_m^*(\v k)P_{mn}(\v q)f_n(\v p),\nn && \Ga_{\v k+\v q,\v
p,\v k,\v p+\v q}\ra \sum_{mn}f_m^*(\v k)C_{mn}(\v q)f_n(\v
p),\nn&& \Ga_{\v k+\v q,\v p,\v p+\v q,\v k}\ra \sum_{mn}f_m^*(\v
k)D_{mn}(\v q)f_n(\v p).\label{projection} \eea Here $\{f_m\}$ is
a set of orthonormal lattice form factors. For honeycomb lattice
the form factor label $m$ also includes a sublattice label,
$m=(m,a)$ with $a=A/B$, within our choice of $C_{3v}$ point group
with respect to the atomic site. (See Appendix.) The decomposition
into each channel would be exact if the set is complete. In
practice, however, a set of a few form factors is often sufficient
to capture the symmetry of the order parameters associated with
leading instabilities.\cite{husemann} The momentum space form
factors are the Fourier transform of the real-space ones: 1)
on-site, $f_{s_0}(\v r)=1$; 2) 1st neighbor bonds, $f_{s_1}(\v
r)=\sqrt{1/3}$, $f_{d_1}(\v r)=\sqrt{2/3}\cos (l\theta_\v r)$ and
$f_{d_{1'}}(\v r)=\sqrt{2/3}\sin (l\theta_\v r)$, where $l=2$ and
$\theta_\v r$ is the azimuthal angle of $\v r$;\cite{note2} 3) 2nd
neighbor bonds, $f_{s_2}(\v r)=\sqrt{1/6}$,
$f_{p_2}=\sqrt{1/3}\cos\theta_\v r$, $f_{p_{2'}}(\v
r)=\sqrt{1/3}\sin\theta_\v r$, $f_{d_2}(\v r)=\sqrt{1/3}\cos
2\theta_\v r$, $f_{d_{2'}}(\v r)=\sqrt{1/3}\sin 2\theta_\v r$,
$f_{f_2}(\v r)=\sqrt{1/6}\cos 3\theta_\v r$. These form factors
(combined with sublattice labels) are used in all channels on
equal footing. We have tested that further neighbor form factors
do not change the results qualitatively.

The one-loop correction to the flow of the vertex function can be
written as, in matrix form, \eqa && \p P/\p \La = P \chi'_{pp} P,
\nn && \p C/\p\La = C \chi'_{ph} C, \nn && \p D/\p\La = (C-D)
\chi_{ph}' D + D \chi_{ph}'(C-D), \label{f1} \eea where the
collective momentum $\v q$ is left implicit for brevity,
$\chi'_{pp/ph}$ are loop integrations projected by the form
factors (See Appendix for details), and $\La$ is the running
cutoff energy. Integrating Eq.~(\ref{f1}) with respect to $\La$
yields the ladder approximation. However, since $\p P$, $\p C$ and
$\p D$ add up to the full change $d \Ga$, the full flow equations
for $P,C$ and $D$ should be given by \eqa && dP/d\La = \p P/\p\La
+ \hat{P} (\p C/\p\La+\p D/\p\La),\nn && dC/d\La =\p C/\p\La +
\hat{C}(\p P/\p\La + \p D/\p\La), \nn && dD/d\La = \p D/\p\La
+\hat{D}(\p P/\p\La + \p C/\p\La), \label{f2} \eea where the
$\hat{P},\hat{C}$ and $\hat{D}$ are the projection operators in
the sense of Eq.~(\ref{projection}), and we used the fact that
$\hat{K}(\p K)=\p K$ for $K=P,C,D$. In Eq.~(\ref{f2}) the terms
preceded by the projection operators represent the overlaps of the
three different channels. It is those terms which allow pairing to
be induced by virtual particle-hole scattering
processes\cite{zhai}.

It can be shown that the effective interaction in the
superconducting (SC), spin density wave (SDW) and charge density
wave (CDW) channels are given by $V_{sc}=-P$, $V_{sdw}=C$, and
$V_{cdw}=C-2D$, respectively. By singular value decomposition, we
determine the leading instability in each channel,\eqa
V_{X}^{mn}(\v q_X)=\sum_\al S_X^\al\phi_X^\al(m)\psi_X^\al(n),\eea
where $X=sc,sdw,cdw$, $S_X^\al$ is the singular value of the
$\al$-th singular mode, $\phi_X^\al$ and $\psi_X^\al$ are the
right and left eigen vectors of $V_X$, respectively. We fix the
phase of the eigen vectors by requiring $\Re
[\sum_m\phi_X^\al(m)\psi_X^\al(m)]>0$ so that $S_X^\al<0$
corresponds to an attractive mode in the X-channel. In the pairing
channel $\v q_{sc}=0$ addresses the Cooper instability. The
pairing function in the sublattice basis can be constructed from
$\phi_{sc}^\al$, and a further unitary transform is needed to get
the pairing function in the band basis. (See the Appendix for
details.) The ordering wave vector in the SDW/CDW channel $\v q=\v
q_{sdw/cdw}$ is chosen at which $|V_{sdw/cdw}(\v q)|$ is maximal.
We note that such a vector has symmetry-related images, and may
change during the FRG flow before settling down to fixed values.
The RG flow is stopped if any of $|P|_{max}$, $|C|_{max}$, or
$|D|_{max}$ becomes roughly 10 times of the bandwidth.\cite{note3}

More technical details can be found in the Appendix.

\begin{figure}
\includegraphics[width=8.5cm]{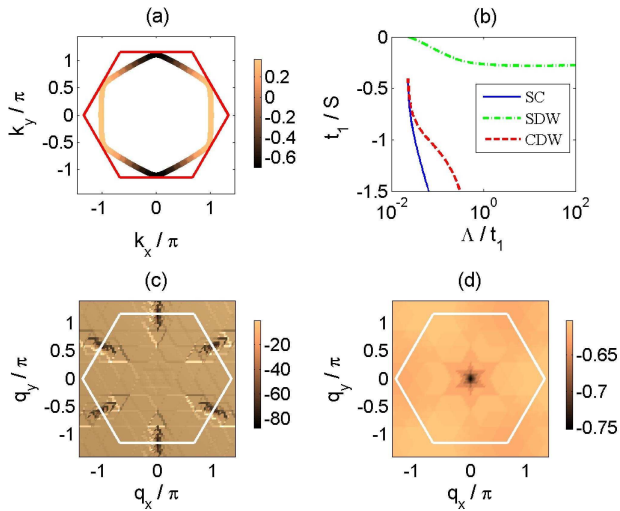}
\caption{(Color online) Results for $\del=1/4$ and $V=0$. (a) The
Fermi surface. The brightness on the surface represents the
momentum space gap function associated with one of the degenerate
pairing modes. (b) FRG flow of (the inverse of) the most negative
singular values $S$ in the SC (blue solid line), SDW (green
dot-dashed line) and CDW (red dashed line) channels. (c) and (d)
are the renormalized interaction $V_{sdw}^{mm}$ for $m=(s_0,A)$,
and $V_{sc}^{mm}$ for $m=(d_1,A)$, as functions of the collective
momentum $\v q$. The hexagons in (a), (c) and (d) indicate the
Brillouine zone boundary.} \label{frg-quarter}
\end{figure}

\section{Results and discussion}

We define the doping level by $\del = n_e-1$ where $n_e$ is the
number of electrons per site. We first discuss the results for
$\del =1/4$ and $V=0$. Fig.\ref{frg-quarter}(a) shows the Fermi
surface in this case, which is well nested and close to the van
Hove singularities (the mid points of edges of the outer hexagon).
The flow of the most negative singular values (denoted as $S$) in
the SC, SDW and CDW channels are shown in
Fig.\ref{frg-quarter}(b). Clearly the SDW (green dot-dashed line)
is the leading instability. This is because the SDW scattering is
already attractive at high energies and is further enhanced by the
Fermi surface nesting shown in Fig.\ref{frg-quarter}(a) down to
the lowest energies. This SDW singular mode $\phi_{sdw}(m)$ has a
dominant value for $m=(s_0,A/B)$, showing that the magnetic
ordering moment is site-local. It is also identifiable in the
renormalized interaction $V_{sdw}^{mm}(\v q)$ for $m=(s_0,A)$
shown in Fig.\ref{frg-quarter}(c), which has strong peaks at three
independent momenta ${\v Q}_1=(0,2\pi/\sqrt{3})$, ${\v
Q}_2=(-\pi,\pi/\sqrt{3})$, ${\v Q}_3=(\pi,\pi/\sqrt{3})$ and their
symmetric images. They define the possible ordering vectors $\v
q_{sdw}$ for the SDW order. In contrast, the scattering associated
with pairing (blue solid line) is initially repulsive in all
channels, and only becomes attractive in the $d$-wave channel
after the SDW scattering grows strong. Regarding Cooper pairing we
find two degenerate leading form factors: the $d_{x^2-y^2}$ and
$d_{xy}$ doublets. One of these is used to generate the momentum
space gap function shown in Fig.~\ref{frg-quarter}(a). The
singular pairing mode $\phi_{sc}(m)$ is nonzero for the 2nd
neighbor bonds, but the amplitude is about one order of magnitude
smaller than that for the 1st neighbor bonds, hence justifying the
cutoff in the real-space range of the form factors. The
renormalized $V_{sc}^{mm}(\v q)$ for $m=(d_1,A)$ in
Fig.\ref{frg-quarter}(d) has negative (but weak) peaks at $\v
q=0$, confirming the Cooper instability at this momentum. The CDW
channel (red dashed line) also becomes weakly attractive from
Fig.\ref{frg-quarter}(b). We checked that the singular mode
$\phi_{cdw}(m)$ has significant values for $m=(s_1,A/B)$ and
$m=(d_{1,1'},A/B)$, showing that it is an extended CDW. The
mixture of $s_1$ and $d_{1,1'}$ is due to the fact that the CDW
wave vector $\v q=(2\pi/3,\pi)$ (or its symmetry images) is not
invariant under the point group operations. However, the CDW
channel remains weak in our case hence will not be discussed in
the rest of the paper. (The merging of SC and CDW lines in
Fig.\ref{frg-quarter}(b) is induced by the diverging SDW channel
via the overlapping among these channels.)

\begin{figure}
\includegraphics[scale=0.32]{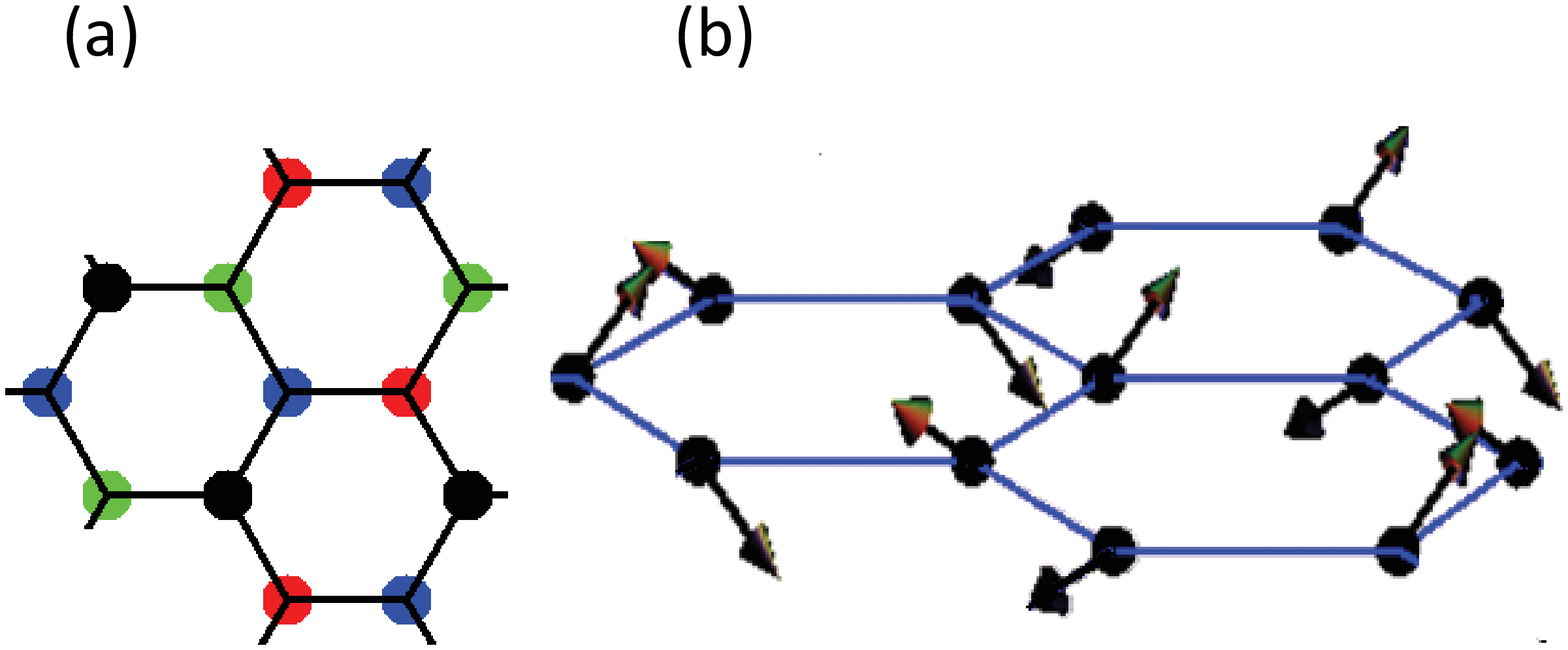} \vspace{-0.7in}
\caption{ (Color online) Chiral SDW order on the honeycomb
lattice. (a) The spins on the black,red,green and blue sublattices
(of different gray scales) are ${\v M}_1+{\v M}_2+{\v M}_3$, $-{\v
M}_1-{\v M}_2+{\v M}_3$, ${\v M}_1-{\v M}_2-{\v M}_3$, $-{\v
M}_1+{\v M}_2-{\v M}_3$ respectively. (b) A 3D perspective view of
the chiral SDW order.} \label{tao-state}
\end{figure}

The above results indicate three independent and degenerate SDW
momenta (apart from the global spin $SU(2)$ symmetry). A
calculation that keeps the symmetry-breaking self-energy flow is
needed to decide which combination of them is realized in the
ordered state, but this is beyond the scope of the present work.
Alternatively, one may resort to a Landau theory or mean field
theory. Indeed, according to the mean field theory of
Ref.~\cite{tao}, a particular linear combination \cite{tao} $
\langle {\v S}_{{\v R},A}\rangle= {\v M}_3 e^{i {\v Q}_3\cdot {\v
R}}+{\v M}_1 e^{i {\v Q}_1\cdot {\v R}} +{\v M}_2 e^{i {\v
Q}_2\cdot {\v R}} $ and $ \langle {\v S}_{{\v R},B}\rangle= {\v
M}_3 e^{i {\v Q}_3\cdot {\v R}}-{\v M}_1 e^{i {\v Q}_1\cdot {\v
R}} -{\v M}_2 e^{i {\v Q}_2\cdot {\v R}} $, gives the most
energetically favorable spin structure. Here ${\v R}$ labels unit
cell and ${\v M}_{1,2,3}$ are three mutually orthogonal and equal
length vectors. The handedness of the ${\v M}_{1,2,3}$ triad
breaks time-reversal and spatial reflection symmetry. The
resulting four-sublattice chiral SDW order is shown in
Fig.~\ref{tao-state}.

\begin{figure}
\includegraphics[width=8cm]{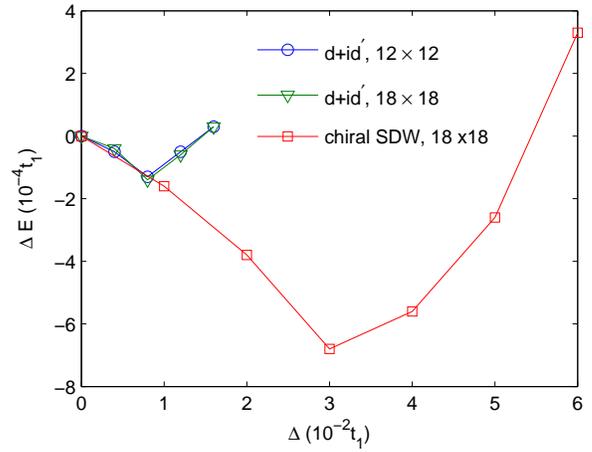}
\caption{(Color online) Variational Monte Carlo results for the
energy gain per site, $\Del E$, versus the variational order
parameters $\Del$ for the $d+id'$-SC (circles and triangles) and
the chiral SDW states (squares) at the doping level $\del=1/4$.
The lattice sizes are given in the legends. Lines are guides to
the eye. \label{vmc}}
\end{figure}

Since our result is at odd with that of Ref.~\cite{chubukov} which
applies in the limit of vanishing interaction strength, we further
check the above conclusion by a variational Monte-Carlo (VMC)
calculation using exactly the same parameters as for
Figs.\ref{frg-quarter}. We adopted the partially-projected
mean-field wave-functions\cite{Fan} as our trial wave-functions,
with the the form factors guided by the present SM-FRG result.
Fig.~\ref{vmc} shows the energy gain per site due to $d+id'$
pairing on $12\times 12$ (circles) and $18\times 18$ (triangles)
lattices, showing negligible size dependence. We then compare to
the energy gain associated with the chiral SDW (squares). It is
clear that the SDW state is far more energetically favorable than
the chiral $d+id'$-SC state at this doping level. The reason that
the SDW state is realized in our lattice model lies in the fact
that the perfect Fermi surface nesting is as important as the
inter-saddle scattering addressed in Ref.\cite{chubukov}.

\begin{figure}
\includegraphics[width=8.3cm]{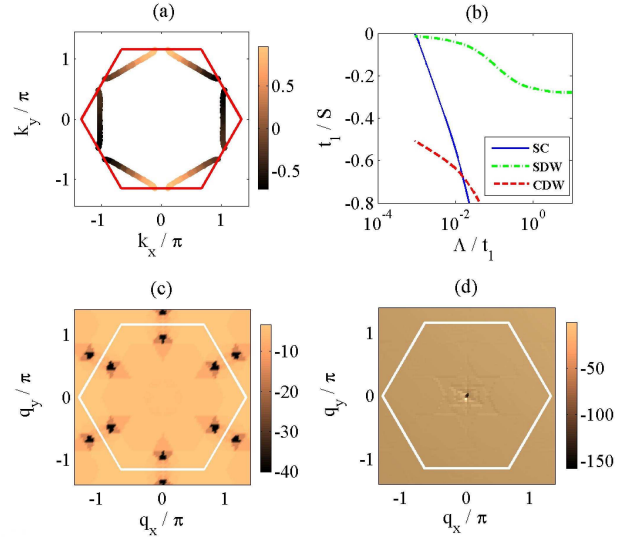}
\caption{(Color online) The same plots as in
Fig.~\ref{frg-quarter} but for $\del=0.211$. Note the splitting of
the SDW peaks in panel (c) signifies the incommensurate SDW
instability.} \label{frg-away}
\end{figure}

Below 1/4 doping, the the Fermi level moves away from the van Hove
points and the Fermi surface nesting worsens. This is shown in
Fig.~\ref{frg-away}(a) for $\del=0.211$ as an example. The bare
interactions are still set as $U=3.6t_1$ and $V=0$. The SDW
scattering is still attractive at high energies. As seen in
Eq.(\ref{f2}), this relatively strong SDW channel causes
attraction in the SC channel via overlap between these channels
(terms with the projection operator in \Eq{f2}). At even lower
energy scales the pairing channel attraction (with $\v q=0$)
continues to grow due to the Cooper instability, while the
enhancement of the SDW scattering is saturated due to the lack of
Fermi surface nesting. As the result the pairing instability
surpasses the SDW instability at the lowest energy scale. This is
shown by the flow of the singular values in
Fig.~\ref{frg-away}(b). It is worth to mention that precisely the
same phenomenon was observed in the FRG studies of the cuprates
and pnictides\cite{honerkamp,zhai}. A close inspection of the
eigenvectors $\phi_{sc}(m)$ associated with the most diverging
superconducting pairing channel again find the degenerate
$d_{x^2-y^2}$ and $d_{xy}$ doublets, with dominant amplitudes for
$m=(d_{1,1'},A/B)$. The momentum space gap function of one of the
pairing modes is shown in Fig.~\ref{frg-away}(a).
Fig.~\ref{frg-away}(c) shows the renormalized $V_{sdw}^{mm}(\v q)$
for $m=(s_0,A)$, which shows weak peaks at six independent and
incommensurate momenta. Fig.~\ref{frg-away}(d) shows the
renormalized $V_{sc}^{mm}(\v q)$ for $m=(d_1,A)$, which shows a
strong negative peak at $\v q=0$.

The above results imply degenerate $d$-wave pairing instabilities.
As for the degenerate SDW instabilities, additional analysis, such
as the mean field theory or Ginzburg-Landau theory, is needed to
fix the structure of the pairing function in the ordered state. To
a large extent, this kind of analysis has been performed in
Ref.\cite{chubukov}. We have also performed simple mean field
calculations using the renormalized pairing interaction. The
result is that a time-reversal breaking $d_{x^2-y^2}\pm
id_{xy}$-wave pairing is always more favorable. This could have
been anticipated since both $d_{x^2-y^2}$ and $d_{xy}$ form
factors have nodes on the Fermi surface, a natural way to gain
energy is to form the above chiral d-wave pairing, which gaps out
the entire Fermi surface.

\begin{figure}
\includegraphics[width=8.5cm]{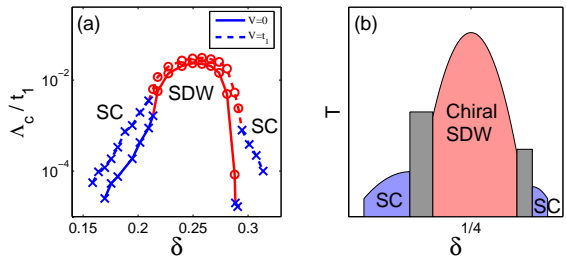}
\caption{(Color online) (a) The FRG diverging energy scale $\La_c$
plotted as a function of doping level near $\delta=1/4$. crosses
and open circles represent $\La_c$ associated with the SC and SDW
channel, respectively. $V=0,t_1$ for solid and dashed lines,
respectively. (b) A schematic temperature-doping phase diagram
near $\del=1/4$ in linear scales. The grey region denotes the
transition between SC and SDW.} \label{phase-diagram}
\end{figure}

We have also checked nearby doping and analyzed the competition
between the incommensurate SDW and the SC state. In
Fig.~\ref{phase-diagram}(a) we plot the higher diverging scale
among these two competing orders as a function of doping (solid
line). We found a similar phase diagram near -1/4 (hole) doping
(not shown), mirroring that of electron doping. (Notice that the
particle-hole symmetry is not exact in the presence of hopping
integrals $t_{2,3}$).

We now discuss briefly the effect of the nearest neighbor
interaction $V$. As a typical example, we set $U=3.6t_1$ and
$V=t_1$, and perform the FRG calculations. We find that the
results are qualitatively similar to the cases with $V=0$, except
that in the leading pairing singular mode,
$\phi_{sc}(d_{2,2'},A/B)$ becomes slightly stronger, but still
smaller than $\phi_{sc}(d_{1,1'},A/B)$ by a factor of $4\sim 6$.
The phase diagram for $V=t_1$ is also drawn in
Fig.\ref{phase-diagram}(a) (dashed line). The critical scale is
slightly higher than the case of $V=0$. In the SC region this is
due to the slight enhancement of $d$-wave pairing on 2nd neighbor
bonds. Unlike that claimed in Ref.\cite{ronny}, in all cases
studied in this paper the $f$-wave pairing is not a leading
instability.

We end by presenting Fig.~\ref{phase-diagram}(b) as a schematic
phase diagram in the temperature-doping plane. In reality when the
doping level slightly deviates from 1/4, the extra charges will be
localized by the presence of disorder, which enables the system to
stay in the Chern insulator state for a finite doping interval. In
the transition region marked by gray, where the doped charges
delocalize, incommensurate SDW states with wave vectors shown in
Fig.\ref{frg-away}(c) will emerge.

\section{Summary}

In summary, we have performed SM-FRG calculations for parameters
suitable for graphene. Our results indicate that near 1/4
electron- or hole- doping, graphene is either a Chern insulator or
a chiral d-wave superconductors. Both phases are topological in
nature, and deserve experimental efforts in searching them.

\acknowledgments{ We thank Hong Yao for helpful discussions, and
are grateful to Tao Xiang for computing resources. QHW
acknowledges the support by NSFC (under grant No.10974086 and
10734120) and the Ministry of Science and Technology of China
(under grant No.2011CBA00108 and 2011CB922101). FY acknowledges
the support by NSFC (under grant No.10704008). DHL acknowledges
the support by the DOE grant number DE-AC02-05CH11231.}

\section{Appendix}

To illustrate the idea of the method, we first ignore the
sublattice index, and return to it at a later stage.

\begin{figure}
\includegraphics[width=8cm]{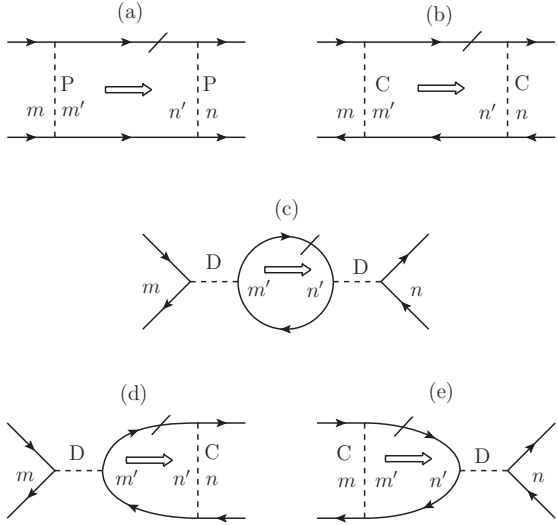}
\caption{One loop diagrams contributing to the flow of the the
4-point vertex function in the pairing channel (a), crossing
channel (b), and direct channel (c)-(e). Here $m,m'n,n'$ denote
form factors, while the momentum and spin indices are left
implicit. The open arrows indicate the flow of the collective
momentum. The slashed lines are single-scale fermion propagators.
The slash can be placed on either internal lines associated with
the loop.}\label{oneloop}
\end{figure}

The one-loop contributions to the flow of the irreducible 4-point
vertex function is shown in Fig.\ref{oneloop}, where (a) and (b)
lead to the flow of $P$ and $C$, respectively, and (c)-(e) lead to
the flow of $D$. The internal Greens functions are convoluted with
the form factors, hence
\begin{widetext} \eqa
(\chi'_{pp})_{mn}&&=\frac{\p}{\p\La}\int\frac{d\w_n}{2\pi}\int\frac{d^2\v
p}{S_{BZ}}f_m(\v p)G(\v p+\v q,i\w_n)G(-\v p,-i\w_n)f_n^*(\v
p)\theta(|\w_n|-\La) \nn &&=-\frac{1}{2\pi}\int\frac{d^2\v
p}{S_{BZ}}f_m(\v p)G(\v p+\v q,i\La)G(-\v p,-i\La)f_n^*(\v p)\ \
+(\La\ra -\La),\nn
(\chi'_{ph})_{mn}&&=\frac{\p}{\p\La}\int\frac{d\w_n}{2\pi}\int\frac{d^2\v
p}{S_{BZ}}f_m(\v p)G(\v p+\v q,i\w_n)G(\v p,i\w_n)f_n^*(\v
p)\theta(|\w_n|-\La) \nn &&=-\frac{1}{2\pi}\int\frac{d^2\v
p}{S_{BZ}}f_m(\v p)G(\v p+\v q,i\La)G(\v p,i\La)f_n^*(\v p)\ \
+(\La\ra -\La),\label{loopint} \eea \end{widetext} where $G$ is
the bare fermion propagator, $S_{BZ}$ is the area of the
Brillouine zone. Here $\La>0$ is the infrared cutoff of the
Matsubara frequency $\w_n$.\cite{karrasch} As in usual FRG
implementation, the self energy correction and frequency
dependence of the vertex function are ignored.

In general, the form factor $f_m(\v k)=\sum_{\v r} f_m(\v r)
\exp(-i\v k\cdot\v r)$, where $f_m(\v r)$ transforms according to
an irreducible representation of the point group, and $\v r$ is
the relative position vector between the two fermion fields on
each side of the diagrams in Fig.\ref{pcd} (b)-(c). For two types
of diagrams to overlap, all of the four fermion fields sit within
the range set by the form factors. Hence the projections in
Eq.(\ref{f2}) are all preformed in real space.

We now return to the honeycomb lattice with two sublattices. The
necessary modifications are as follows: 1) The sublattice label
can be absorbed into the labels $1,2,3,4$ in Figs.\ref{pcd}, so in
principle the label $m$ in the form factor $f_m(\v r)$ also
includes sublattice indices. However, once $\v r$ is fixed they
are not independent. We absorb an independent index into the form
factor label, $m\ra (m,a)$, where $a$ labels fermion field 1 or 4
in Fig.\ref{pcd}(b), 1 or 4 in (c), and 1 or 3 in (d). This is
reasonable because the point group operations do not mix the
sublattice when the center is chosen to be atomic sites. 2) In the
presence of sublattice indices the Green's functions are matrices.
3) In order to ensure that in momentum space $P$, $C$ and $D$
transform exactly as product of form factors, care must be taken
in choosing the phase of the Bloch states for complex unit cells.

The pairing function is determined as follows. A singular mode
$\phi_{sc}^\al$ corresponds to a pairing operator $\sum_{m=(m,a)}
c_{a\ua}^\dag (\v k)\phi_{sc}^\al(m)f_m(\v k)^*c_{a_m\da}^\dag
(-\v k)$ in the momentum space, where $a_m$ is determined by
$m=(m,a)$ (for all associated vectors $\v r$). The parity of the
pairing matrix function under space inversion determines
automatically whether it is a spin singlet or spin triplet. A
further unitary transform can be used to get the momentum space
gap function.

\begin{figure}
\includegraphics[width=8.5cm]{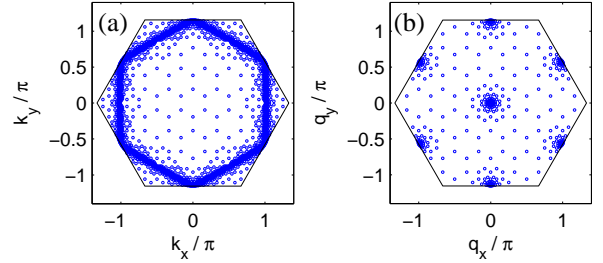}
\caption{(Color online) Examples of self-adaptive $\v k$-mesh (a)
and $\v q$-mesh (b) used for loop integrations and collective
momenta for interactions, respectively. Notice that in (a) the
mesh is too dense near the Fermi surface to be differentiated by
naked eyes.} \label{mesh}
\end{figure}

In the current implementation of SM-FRG the sampling of momentum
space ($\v k$ and $\v q$) is performed on self-adaptive meshes. As
an example, Fig.\ref{mesh}(a) shows the $\v k$-mesh, which is
progressively denser in approaching the Fermi surface or van Hove
points (if they are close to the Fermi level). This is important
for our problem because of the rapid variation of the Fermi
velocity near the van Hove singularities. The $\v k$-mesh is
obtained as follows. First define an energy scale $\Omega$ (of the
order of the bandwidth), and begin with 6 equal-area triangles
spanning the BZ. Break a specific triangle into four smaller
equal-area ones if any eigen energy $|\eps_\v k|\leq \Omega$ in
the original triangle. Then lower the energy scale as $\Omega\ra
\Omega/b$ ($b>1$) and repeat the above process recursively. The
centers of the triangles form the mesh points. Combining the
triangle areas, they are used in the loop integrations in
Eq.(\ref{loopint}). In our implementation, we perform the above
processes 8-10 times, so that the last generation of triangles has
a linear size of order $2^{-8}\pi\sim 2^{-10}\pi$, and the center
of such triangles are sufficiently close to (or accidentally on)
the Fermi surface. The $\v q$-mesh as in Fig.\ref{mesh}(b) used
for the interactions includes all important scattering momenta:
the origin and the high symmetry nesting vectors. We devise a
function $\eta_\v q$ such that it is zero at those important
scattering momenta, and generate the mesh in a similar fashion as
for $\v k$, except that $\Omega$ becomes an artificial scale and
$\eta_\v q$ is used in place of $\eps_\v k$. Similar q-mesh
already appears in Ref.\cite{husemann}.


\begin{references}
\bibitem{note2} We note by setting $l=1$ above
does not produce independent spatial form factors because the
above three form factors are complete for the three 1st neighbor
bonds. Hence by using $f_{d_1,d_{1'}}$ we are able to investigate,
e.g., both $p$- and $d$-wave pairing (of course their total form
factors have completely different spin and sublattice structures).
\bibitem{note3} Translating $P$, $C$ and $D$ back to $\Ga_{1234}$, the divergence
criterion is smaller since two form factors must be accounted for.
\end{references}
\end{document}